\documentclass[aps,prb,twocolumn]{revtex4-1}

\usepackage{amsmath}
\usepackage{graphicx}
\usepackage{dcolumn}
\usepackage{bm}
\usepackage{mathrsfs}
\usepackage{mhchem}
\usepackage{subfigure}
\usepackage{color}

\draft

\begin{document}

\title{Chemotactic and hydrodynamic effects on collective dynamics of self-diffusiophoretic Janus motors}

 \author{Mu-Jie Huang}
\email{mjhuang@chem.utoronto.ca}
\author{Jeremy Schofield}
\email{jmschofi@chem.utoronto.ca}
\author{Raymond Kapral}
 \email{rkapral@chem.utoronto.ca}
 \affiliation{Chemical Physics Theory Group, Department of Chemistry, University of Toronto, Toronto, Ontario M5S 3H6, Canada}

\date{\today}

\begin{abstract}
Collective motion in nonequilibrium steady state suspensions of self-propelled Janus motors driven by chemical reactions can arise due to interactions coming from direct intermolecular forces, hydrodynamic flow effects, or chemotactic effects mediated by chemical gradients. The relative importance of these interactions depends on the reactive characteristics of the motors, the way in which the system is maintained in a steady state, and properties of the suspension, such as the volume fraction. From simulations of a microscopic hard collision model for the interaction of fluid particles with the Janus motor we show that dynamic cluster states exist and determine the interaction mechanisms that are responsible for their formation. The relative importance of chemotactic and hydrodynamic effects is identified by considering a microscopic model in which chemotactic effects are turned off while the full hydrodynamic interactions are retained.  The system is maintained in a steady state by means of a bulk reaction in which product particles are reconverted into fuel particles.  The influence of the bulk reaction rate on the collective dynamics is also studied.

\end{abstract}

%
%
%
%
%

\maketitle

\section{Introduction}

A large number of molecular machines have evolved in the natural world that are able to operate singly and collectively to perform essential biological functions at the nanoscale~\cite{alberts-cell}. On somewhat larger scales bacteria and other swimming microorganisms behave collectively by performing work on the fluid in which they are immersed, and it has been shown that effects arising from hydrodynamic interactions among such active swimmers are of crucial importance to understand the origins and characteristics of their collective motion~\cite{lauga2009,Baskaran_Marchetti_09,saintillan2012,Spagnolie_Lauga_12,Marchetti_etal_13,Bricard_etal_13,Zottl_Stark_14,Blaschke_etal_16}.
More generally, a considerable research effort has been devoted to the study of the collective dynamics observed in various types of media that contain actively propelled particles, due, in part, to the rich nonequilibrium phenomena they exhibit~\cite{vicsek1995,chate2008,peruani2006,redner2013,cates2013,bialke2013,Palacci_eta_13,wysocki2014,takatori2015,speck2015,Zottl_Stark_16}.
Among such active particles are synthetic chemically-propelled nanomotors, which have also been studied extensively~\cite{SenRev:13,wangbook:13,kapral2013,Ma_Hahn_Sanchez_15}. Synthetic nanomotors are of particular interest since their morphology and chemical properties may be tailored for specific tasks~\cite{Jiang_etal_10}. Ensembles of such chemically-powered motors display collective behavior that is influenced by the chemical gradients that exist in these systems, as well as hydrodynamic flow effects~\cite{ibele2009,Theurkauff_etal_12,Thakur_Kapral_12,Wang_etal_13,Buttinoni_etal_13,kapral2014,Pohl_Stark_14,Saha_etal_14,Pohl_Stark_15,wang2015,huang_kapral_16,Colberg_Kapral_17}.

In this paper we study the collective properties of suspensions of spherical Janus motors whose activity arises from a diffusiophoretic mechanism. In the self-diffusiophoretic mechanism, asymmetric catalytic activity on the surface of a nanomotor produces inhomogeneous local concentrations of chemical species that give rise to  a body force on the motor that is responsible for its propulsion~\cite{anderson1989,golestanian2005,kapral2013}. As a result of momentum conservation, a self-propelled motor moving in a fluid medium is coupled to the motions of solvent molecules and generates fluid flows which are an essential element in the diffusiophoretic mechanism~\cite{anderson1989}. Experimental studies of suspensions of Janus motors have shown dynamical clustering, phase separation and giant number fluctuations~\cite{Palacci_eta_13,Theurkauff_etal_12,Buttinoni_etal_13}. Chemotactic effects that arise from concentration-mediated interactions among motors have been shown to play a role in the collective dynamics of such chemically-powered motors~\cite{Theurkauff_etal_12,Pohl_Stark_14,Saha_etal_14,Pohl_Stark_15}.

The strength and direction of chemotactic and hydrodynamic interactions among motors in a suspension depend on their relative orientations since concentration and flow fields are induced by catalytic reactions taking place on specific parts of the motor surfaces; consequently, motor morphology plays a role in determining the dynamical properties of active suspensions.  The effects of the size of the catalytic patch on the dynamics of a single Janus motor in bulk~\cite{Popescu_etal_10} or near a wall~\cite{Popsescu_etal_16} and on the interactions between two motors~\cite{Mood_etal_09} have been studied using continuum theory.

We investigate the effect of the size of a Janus motor's catalytic surface on the induced flows in the surrounding fluid and on the roles that hydrodynamic and chemical interactions play in the collective behavior of a suspension of Janus motors. Our study is carried out using particle-based simulations employing a microscopic model of hard collisions between motor and fluid particles~\cite{Huang_etal_16}. Modifications of the model are made to isolate the various factors that contribute to collective behavior of active suspensions. The outline of the paper is as follows: In Sec.~\ref{sec:micro_model} a brief review of the microscopic model is presented, followed by a discussion of the steady-state concentration and flow fields for a single Janus motor with various catalytic cap sizes. In Sec.~\ref{sec:collective} the collective properties of systems containing many Janus motors are examined, and the structural features and dynamical behavior of suspensions of different types of motors as a function of volume fraction and steady state conditions are discussed with particular attention to the relative importance of chemotactic versus hydrodynamic interactions among motors. A summary of results is given in Sec.~\ref{sec:conclusion}.

\section{Janus motor and its nonequilibrium dynamics}\label{sec:micro_model}

It is useful to discuss some of the features of a single Janus motor and its dynamics before describing the collective dynamics of many motors. We consider a hard-sphere model of a chemically-powered self-propelled Janus motor moving in a fluid medium containing $N_A$ fuel $A$ and $N_B$ product $B$ particles~\cite{Huang_etal_16}. Irreversible chemical reactions, $A \to B$, take place on the catalytic portion of the Janus motor surface and are the source of concentration gradients that lead to diffusiophoretic motor propulsion. The system is maintained in a nonequilibrium steady state through irreversible chemical reactions, $B \stackrel{k_2}{\rightarrow} A$, in the bulk of the solution, and the rate at which these bulk phase reactions occur controls the magnitudes of the steady state concentration fields. The bulk reaction rate also determines how rapidly these fields decay with distance from the surface of the Janus motor through their dependence on the inverse screening length $\kappa= \sqrt{k_2/D}$, where $D$ is the common $A$ and $B$ species diffusion coefficient. The evolution of the entire system is governed by a coarse-grain dynamics that combines molecular dynamics and multiparticle collision dynamics~\cite{Malevanets_Kapral_99,Malevanets_Kapral_00,kapral:08,gompper:09} and is described in Appendix~\ref{app:sim}.

\begin{figure}[htbp]
\centering
\resizebox{0.4\columnwidth}{!}{%
\includegraphics{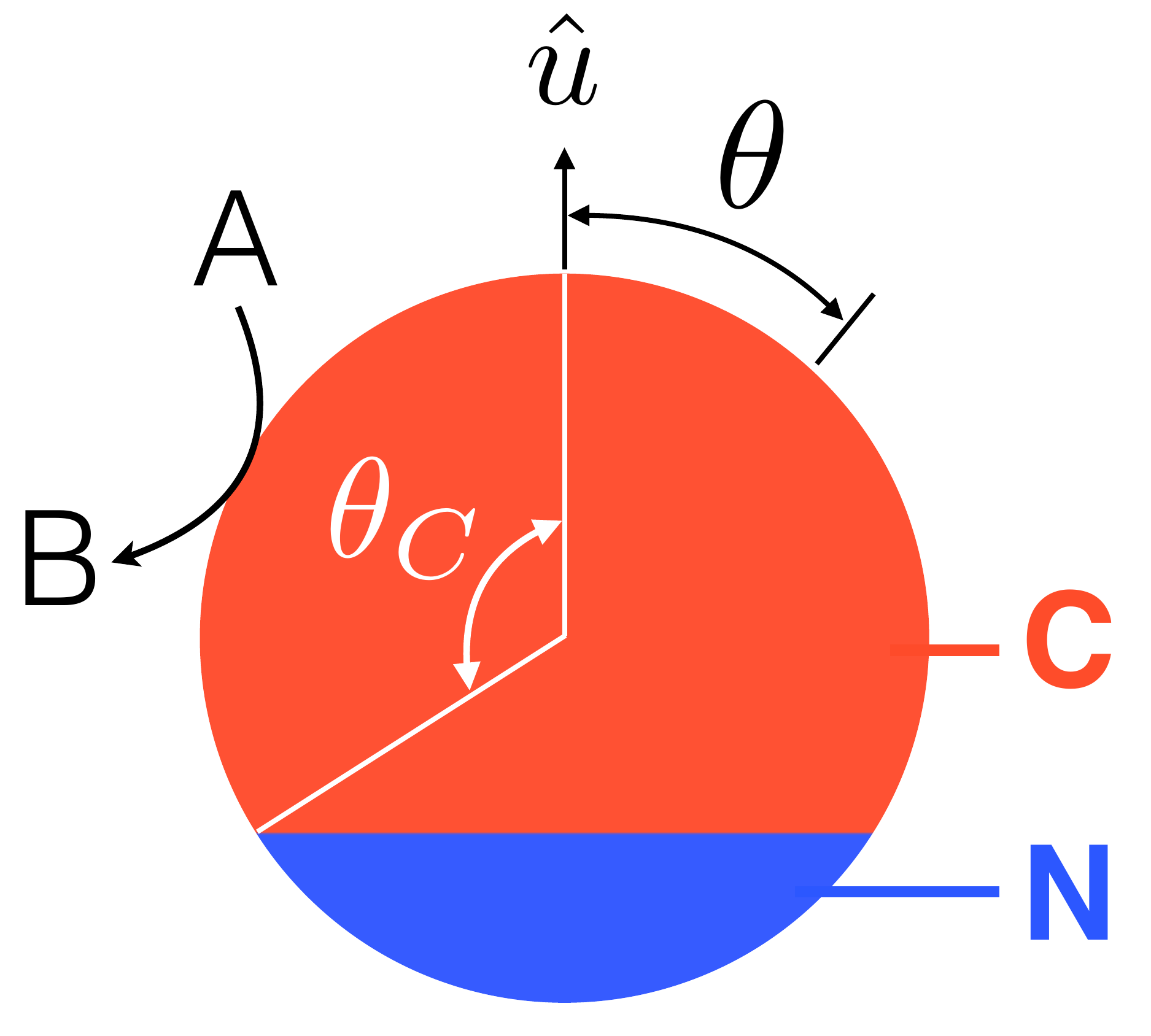}}
\caption{\label{fig:Janus} The Janus motor has catalytic ($C$) and a noncatalytic ($N$) surfaces. The size of the catalytic surface is characterized by the angle $\theta_C$. Chemical reactions, $C+A \to C+B$, occur on the $C$ surface. The unit vector $\hat{{\bf u}}$ is the orientation of the Janus motor in the direction from the $N$ surface to the $C$ surface and $\theta$ is the polar angle.}
\end{figure}
A schematic picture of a Janus motor is shown in Fig.~\ref{fig:Janus}. It is a sphere with catalytic ($C$) and noncatalytic ($N$) surfaces that interacts with the $\alpha = A,B$ species through hard potentials, $W_{\alpha}(r)$, where $W_{\alpha}(r) = \infty$ if $r \leq R_{\alpha}$ and zero when $r > R_{\alpha}$. When a particle of species $\alpha$ encounters the Janus sphere at radius $R_\alpha$ it undergoes a modified bounce-back collision that exchanges linear and angular momenta with the Janus motor~\cite{Huang_etal_16}. The effective radius $R$ of the outer edge of the boundary layer of the Janus motor is chosen as the larger of the two interaction radii. The area of the catalytic surface is determined by the angle $\theta_C$ measured from the pole of the catalytic surface, located by the unit vector $\hat{{\bf u}}$, to the interface between the $C$ and $N$ surfaces. In general reactive collisions, $C+A \to C+B$, occur with probability $p_R$ whenever an $A$ particle encounters the motor catalytic surface ($\theta < \theta_C$), but we choose unit probability $p_R=1$ in this study.

Reactions on the catalytic motor surface create an inhomogeneous concentration field $c_B(\mathbf{r})$. Since the $A$ and $B$ particles have different interaction radii, a body force acts on the motor and is responsible for its propulsion. Because the system is force free, the forces on the motor induce a flow field $\mathbf{v}(\mathbf{r})$ in the surrounding fluid, which is important for the motor self-propulsion mechanism~\cite{anderson:83,Reigh_etal_16}. The continuum approximations to the concentration and velocity fields can be found by solving the coupled reaction-diffusion and Stokes equations. The velocity field at the outer edge of the boundary layer near the motor surface ($r= R$) is given by $\mathbf{v} =\mathbf{v}^{(s)} +  V_u \hat{\mathbf{u}}$, where the slip velocity is~\cite{Julicher_Prost_09}
\begin{equation}
\mathbf{v}^{(s)}(R,\theta) = - \frac{k_BT}{\eta}\Lambda \boldsymbol{\nabla}_{\theta} c_B(R,\theta),
\label{eq:vs}
\end{equation}
for our binary mixture of mechanically identical $A$ and $B$ particles whose specific volumes are equal. In Eq.~(\ref{eq:vs}), the fluid viscosity is $\eta$ and the component of the Janus motor velocity in the $\hat{\mathbf{u}}$ direction is given by $V_u = -\langle \hat{\mathbf{u}} \cdot \mathbf{v}^{(s)} \rangle_S$, where $\langle \cdots \rangle_S$ denotes the surface average at $r = R$. Here the constant $\Lambda =\frac{1}{2}(R_A^2- R_B^2) $ is determined by the different interactions of the $A$ and $B$ species with the Janus motor, as reflected in their different interaction radii, $R_A$ and $R_B $. Due to the fact that catalytic reactions occur only on the $C$ surface, the gradient field, $\boldsymbol{\nabla}_{\theta}c_B$, changes rapidly across the interface between the $C$ and $N$ surfaces along the $\theta$ direction. Therefore, one expects that the largest slip velocity will occur at $C-N$ interface and $|\mathbf{v}^{(s)}(R,\theta_C)| > |V_u|$.

\begin{figure*}[htbp]
\centering
\resizebox{1.7\columnwidth}{!}{%
\includegraphics{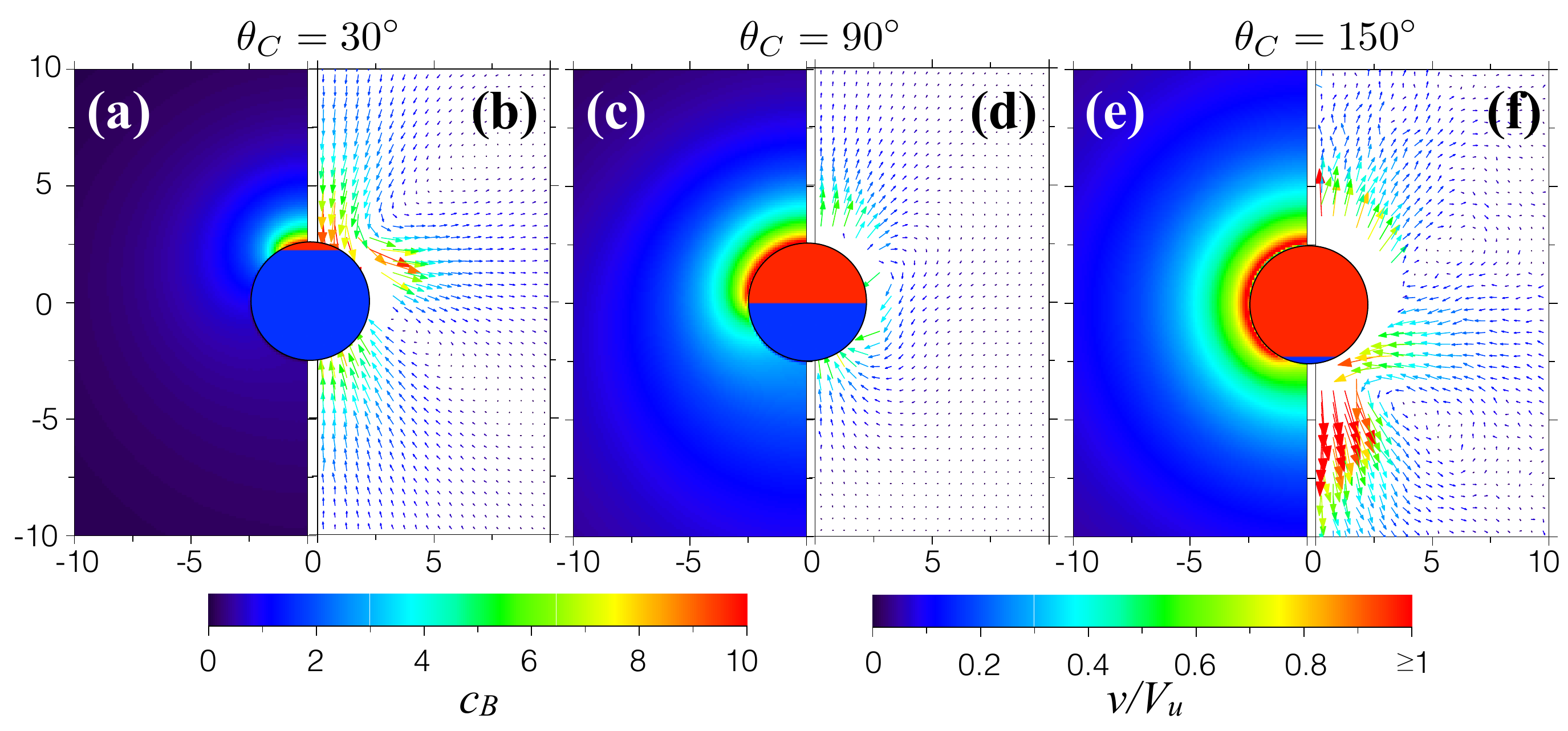}}
\caption{\label{fig:cB_solvent_flow} Panels (a), (c) and (e) are the simulation results for the steady-state product concentration fields ($c_B$), and panels (b), (d) and (f) are the induced flow field ($v$) normalized by the propulsion speed ($V_u$) for three different sizes of the catalytic surface: $\theta_C = 30 ^{\circ}$ (left), $90^{\circ}$ (middle) and $150^{\circ}$ (right). The interaction radii are $R_A = 2.5$ and $R_B = 2.45$ and bulk reaction rate constant is $k_2 = 10^{-3}$.}
\end{figure*}

The profiles of the concentration and velocity fields vary with the surface area of catalytic cap and the inverse screening length $\kappa$. We first consider fixed interaction radii $R_A = 2.5$ and $R_B = 2.45$, and $\kappa\simeq 0.12$. Since $R_A > R_B$, $\Lambda > 0$, and the Janus motor interacts more weakly with the $B$ particles than with $A$ particles giving rise the motor self propulsion in the forward direction (forward-moving motor) defined as being in the direction of $\hat{\mathbf{u}}$ (see Fig.~\ref{fig:Janus}). The catalytic cap areas we consider are derived from $\theta_C= 30^{\circ}$, $90^{\circ}$ and $150^{\circ}$. Figure~\ref{fig:cB_solvent_flow} compares the steady-state product concentration, $c_B(\mathbf{r})$, and flow fields, $\mathbf{v}(\mathbf{r})$, in the vicinity of the Janus motor for these different cap sizes. As expected, the maximum product concentration occurs at the catalytic surface and deceases in both radial and tangential directions from this surface. As noted above, the direction of motor motion is determined by the interaction potentials between the Janus motor and the fuel and product particles. For the forward-moving Janus motors considered here, the near-field flows generated in the surrounding fluid change significantly as the catalytic surface area varies. For the Janus motor with $\theta_C = 30^{\circ}$, there are two incoming flows along the motor axis $\hat{{\bf u}}$ and an outgoing flow in the lateral direction , while the fluid velocity flow fields are in the opposite direction for the Janus motor with $\theta_C = 150^{\circ}$. For $\theta_C = 90^{\circ}$, solvent particles are pushed away in front of the Janus motor and an incoming flow is induced at the rear of the motor, along with a more complicated fluid circulation in the lateral directions~\cite{Reigh_etal_16}.

Making use of the continuum expression for the slip velocity in Eq.~(\ref{eq:vs}), the characteristics in the fluid flow field observed in the simulations can be related to the location of the $C-N$ interface on the motor surface. For the Janus motor with $\theta_C = 30^{\circ}$, this interface lies near the head of the sphere, indicated by the $\hat{\mathbf{u}}$ vector and termed the north pole, inducing a velocity field $\mathbf{v} \simeq \mathbf{v}^{(s)}$ that moves solvent particles from the front of the motor to the lateral directions; the solvent flow field at the south pole is $\mathbf{v} \simeq V_u \hat{\mathbf{u}}$ since the gradient field $\boldsymbol{\nabla}_{\theta}c_B$ is small. Similarly, for a motor with a large catalytic surface ($\theta_C = 150^{\circ}$), the $C-N$ interface is close to the south pole of the sphere producing a flow field that takes solvent particles from the lateral directions to the noncatalytic $N$ surface; solvent particles in front of the motor are pushed away from the motor. For the Janus motor with $\theta_C = 90^{\circ}$, the gradient field $\boldsymbol{\nabla}_{\theta}c_B$ is small near both the north and south poles so that the induced flow is largely determined by the motion of the Janus motor in these two regions.

\section{Collective dynamics in suspensions of Janus motors}\label{sec:collective}

The dynamics of many Janus motors suspended in reactive fluids is governed by direct interactions between motors as well as fluid-mediated chemotactic and hydrodynamic interactions that lead to collective behavior, such as the formation of dynamic nonequilibrium cluster states.  Chemotactic interactions have their origin in the inhomogeneous concentration fields that arise from motor asymmetric catalytic activity.  This inhomogeneity is sensitive to the local environment each motor experiences, and the gradients in concentration created at the surface of one motor influence both the reaction dynamics and the local body forces that drive active motion in surrounding motors.  At the same time the motion of each motor induces flow fields in the surrounding fluid that couple the dynamics of motors via hydrodynamic interactions.  Depending on microscopic characteristics of the interactions, chemotactic and hydrodynamic interactions can work either cooperatively or against one another to enhance or suppress dynamical clustering.  Both the chemotactic and hydrodynamic interactions depend strongly on the size of the catalytic cap on the spherical Janus motors.

In the simulations described below, we consider a system of $N_J$ Janus motors and $N_S$ fluid molecules of species $A$ and $B$ in a periodic cubical box of side length $L$. The motors interact directly with one another through a repulsive Lennard-Jones potential with length scale $\sigma_J$ so that the effective excluded volume is $V_J' = \pi\sigma_J^3/6$ and the motor volume fraction is $\phi = N_J V_J'/L^3$. The system is maintained out of equilibrium by a supply of fuel through reactions $B \stackrel{k_2}{\rightarrow} A$ in the bulk of the solution. When the interaction radii $R_A$ and $R_B$ are chosen so that $R_A > R_B$, each motor moves preferentially toward regions of high production concentration, and on average motors move forward in the direction of the catalytic cap.
On the other hand, when $R_A < R_B$, motors move backward, away from regions of high product concentration.
In previous studies~\cite{Colberg_Kapral_17,Huang_etal_16} it was found that cooperative motion in collections of both backward-moving sphere-dimer and Janus motors was significantly reduced compared to that in systems of forward-moving motors.  For this reason, below we focus on the more interesting clustering properties found in suspensions of forward-moving motors. Unless stated otherwise, for most of the results in this section the interaction radii of the $A$ and $B$ particles are set to $R_A = 2.5$ and $R_B = 2.35$, while the bulk reaction rate constant is chosen to be $k_2 = 10^{-3}$.  We consider systems with $N_J = 100$, $200$ and $500$ Janus motors, corresponding to the volume fractions $\phi = 0.052$, $0.1$ and $0.26$. Further simulation details are given in Appendix~\ref{app:sim}

The mean values of the steady state concentrations of the $A$ and $B$ species depend on the catalytic cap size, motor volume fraction and the rate at which fuel is supplied. In Table~\ref{tab:De} the values of the steady state product concentration, $c_B^{ss}$ ($c_A^{ss}=c_0-c_B^{ss}$), where $c_0$ is the total concentration of $A$ and $B$), are given for systems with various cap sizes and volume fractions. The steady state composition will play some role in determining the collective behavior.
\begin{table}[htbp]
\centering
\caption{\label{tab:De} The steady-state concentrations of product particles, $c_B^{ss}$, mean motor velocity, $\overline{\mbox{V}\raisebox{3.0mm}{}}_u$, (taken from the data shown in Fig.~\ref{fig:Vz_MSD_RB_2_35_compare}(a)), and effective diffusion coefficient, $D_e(\phi)$, (from linear fits to long-time mean-square displacement in Fig.~\ref{fig:Vz_MSD_RB_2_35_compare})(b)), as a function of volume fraction and catalytic cap size.}
\begin{tabular}{l|ccc|ccc|ccc}
\hline
$\theta_C$& &$30^{\circ}$&& &$90^{\circ}$& &&$150^{\circ}$ \\
$\phi$  & $0.052$ & $0.1$  & $0.26$  & $0.052$ &$0.1$  &$0.26$  &$0.052$ &$0.1$ &$0.26$ \\\hline
$c_B^{ss}/c_0$  & $0.2$ & $0.29$ & $0.45$ & $0.51$ & $0.67$ & $0.85$ & $0.34$ & $0.58$ & $0.94$\\
$\overline{\mbox{V}\raisebox{3.0mm}{}}_u$ & $0.014$ & $0.008$ & $0.002$ & $0.021$  &$0.014$ &$0.006$ &$0.0009$ &$0.0008$&$0.0005$ \\
$D_e$ & $0.045$ & $0.020$ & $0.004$ & $0.10$  &$0.050$ &$0.012$ &$0.002$ &$0.002$&$0.002$ \\
\hline
\end{tabular}
\end{table}

\begin{figure*}[htbp]
\centering
\resizebox{1.7\columnwidth}{!}{%
\includegraphics{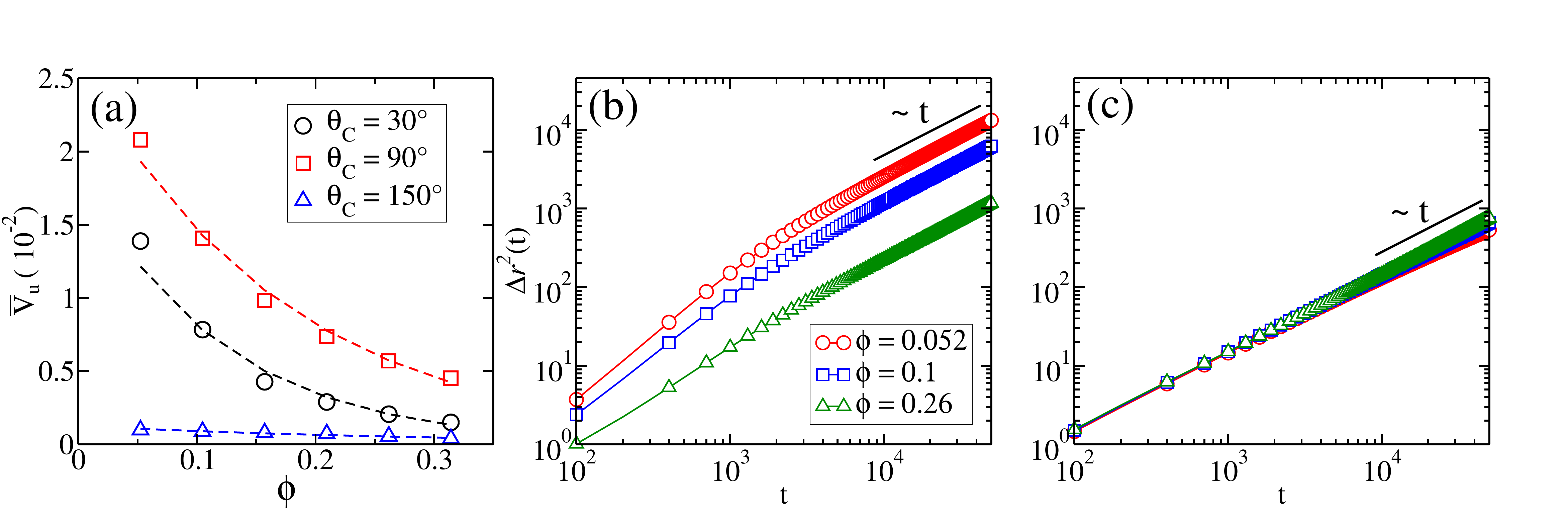}}
\caption{\label{fig:Vz_MSD_RB_2_35_compare} (a) The average motor propulsion velocity, $\overline{\mbox{V}\raisebox{3mm}{}}_u (\phi)$, for the Janus motors with catalytic surface sizes $\theta_C = 30^{\circ}$ (black circles), $90^{\circ}$ (red squares) and $150^{\circ}$ (blue triangles) at various volume fractions ($\phi$), where the dashed lines are the exponential fits to  data as guides to the eye. Mean square displacements, $\Delta r^2(t)$, of the Janus motors with (b) $\theta_C = 30^{\circ}$ and (c) $150^{\circ}$ at various volume fractions. The solid black lines show the asymptotic enhanced diffusive dynamics at long time scales.}
\end{figure*}

Figure~\ref{fig:Vz_MSD_RB_2_35_compare}(a) shows the ensemble and time average in the steady state of the projection of the motor propulsion velocity along $\hat{{\bf u}}$,
\begin{equation}
\overline{\mbox{V}\raisebox{3.0mm}{}}_u (\phi)
 = N_J^{-1} \sum_{i=1}^{N_J} T^{-1} \int_0^T dt\; \mathbf{V}_i(t)\cdot \hat{\mathbf{u}}_i(t)\equiv\langle \mathbf{V}\cdot \hat{\mathbf{u}}\rangle,
\end{equation}
as a function of $\phi$ for motors with different catalytic cap sizes. Here $\mathbf{V}_i(t)$ and $\hat{\mathbf{u}}_i(t)$ denote the instantaneous velocity and orientation of motor $i$.
It is evident that $\overline{\mbox{V}\raisebox{3.0mm}{}}_u (\phi)$ decreases as $\phi$ increases for all cap sizes as might be anticipated for crowded systems. On long time scales, as a result of rotational Brownian motion characterized by the orientational relaxation time, $\tau_r(\phi)$, each of the motors undergoes diffusive translational motion with an effective translational diffusion constant $D_e(\phi)$. This effective diffusion constant is larger than $D_0(\phi)$ for Janus particles without diffusiophoretic activity. It was shown earlier~\cite{Huang_Schofield_Kapral_17} that $D_0(\phi)$ and $\tau_r(\phi)$ do not vary significantly with $\phi$ for $0<\phi \le 0.26$. For a single motor in solution we have $D_0 \approx 0.003$ and $\tau_r \approx 600$. For catalytic cap sizes given by $\theta_C = 30^{\circ}$ and $90^{\circ}$, the general structure of the mean squared displacement $\Delta r^2(t)$ is the same, which consists of a crossover of $\Delta r^2(t)$ from short-time ballistic motion due to self-propulsion to long-time diffusive behavior, as can be seen in Figs.~\ref{fig:Vz_MSD_RB_2_35_compare} (b) for all $\phi$. However, for motors with large catalytic caps
($\theta_C = 150^{\circ}$) $\overline{\mbox{V}\raisebox{3.0mm}{}}_u(\phi)$ is small and the crossover regime cannot be discerned in the figure and $D_e \simeq D_0$.

\begin{figure}[htbp]
\centering
\resizebox{1.0\columnwidth}{!}{%
\includegraphics{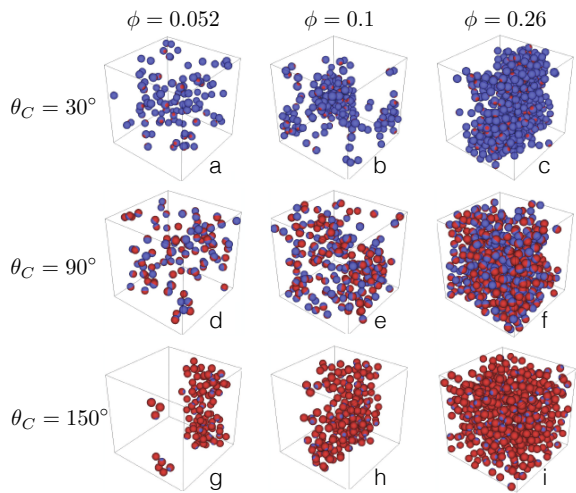}}
\caption{\label{fig:snapshots}Instantaneous steady-state configurations of Janus motors with small ($\theta_C = 30^{\circ}$), medium ($\theta_C = 90^{\circ}$) and large ($\theta_C = 150^{\circ}$) catalytic cap sizes at various volume fractions $\phi = 0.052$, $0.1$ and $0.26$.}
\end{figure}

\subsection{Clustering in suspensions of motors}
\begin{figure*}[htbp]
\centering
\resizebox{1.7\columnwidth}{!}{%
\includegraphics{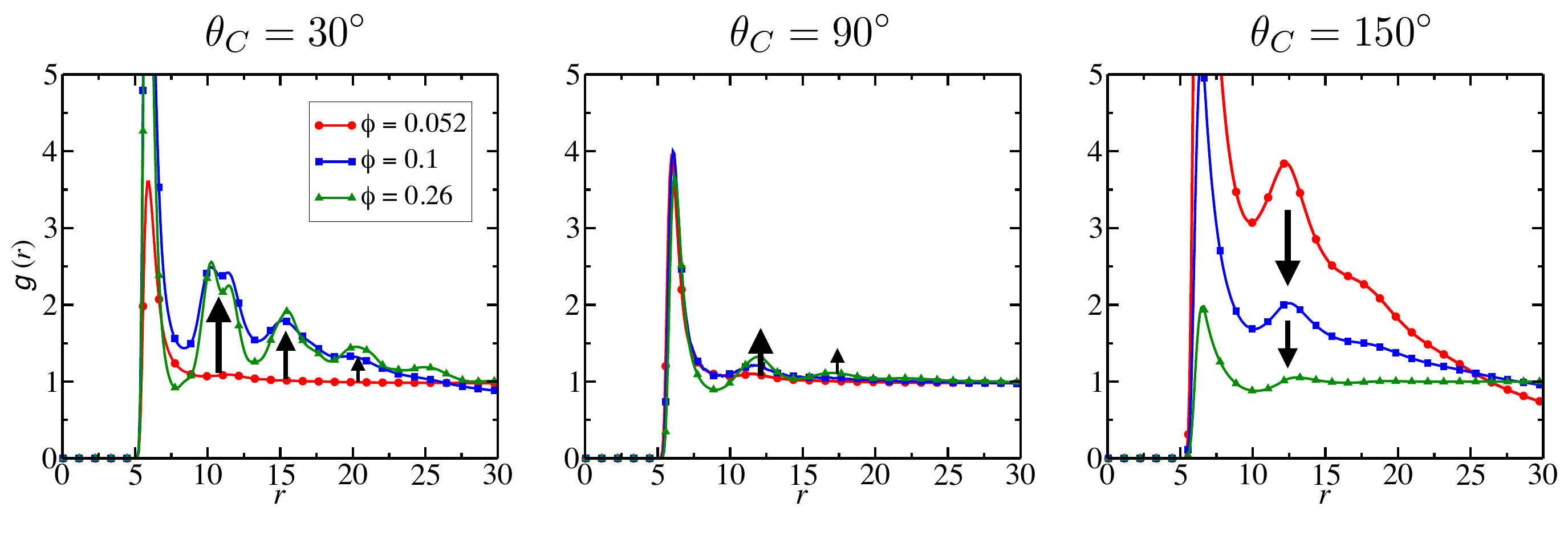}}
\caption{\label{fig:g_r_with_chemo} Radial distribution function, $g(r)$, for the systems of Janus motors with small (left panel), medium (middle panel) and large (right panel) catalytic cap sizes at volume fractions $\phi = 0.052$ (red circles), $0.1$ (blue squares) and $0.26$ (green triangles). The black arrows indicate the changes of $g(r)$ as $\phi$ increases.}
\end{figure*}

For certain parameter values the cooperative dynamics in suspensions of active motors leads to the formation of dynamic clusters.   We now examine some of the properties of these states and factors that lead to their formation. Figure~\ref{fig:snapshots} shows representative instantaneous steady-state configurations of the Janus motors (the solvent species are not shown) in systems with $\phi = 0.052$, $0.1$ and $0.26$. As $\phi$ increases, separation into low-density and high-density regions is observed for motors with a small cap size ($\theta_C = 30^{\circ}$). In contrast, phase separation in systems of Janus motors with a large cap size ($\theta_C = 150^{\circ}$) is observed only at low volume fractions. A weak tendency to cluster is observed for the intermediate cap size ($\theta_C = 90^{\circ}$) as $\phi$ varies. To quantitatively describe motor clustering, we compute the motor radial distribution function in the steady state, $g(r)$, for various volume fractions.  As indicated by the black arrows in the plots of $g(r)$ in Fig.~\ref{fig:g_r_with_chemo}, it is evident that as $\phi$ increases,  clustering is strongly enhanced for the system with small-cap-size motors, a weak enhancement is seen for motors of intermediate cap size, whereas clustering is suppressed for motors with large catalytic caps.

\subsection{Clustering mechanisms for Janus motors with large caps}\label{sec:150}
\begin{figure}[htbp]
\centering
\resizebox{1.0\columnwidth}{!}{%
\includegraphics{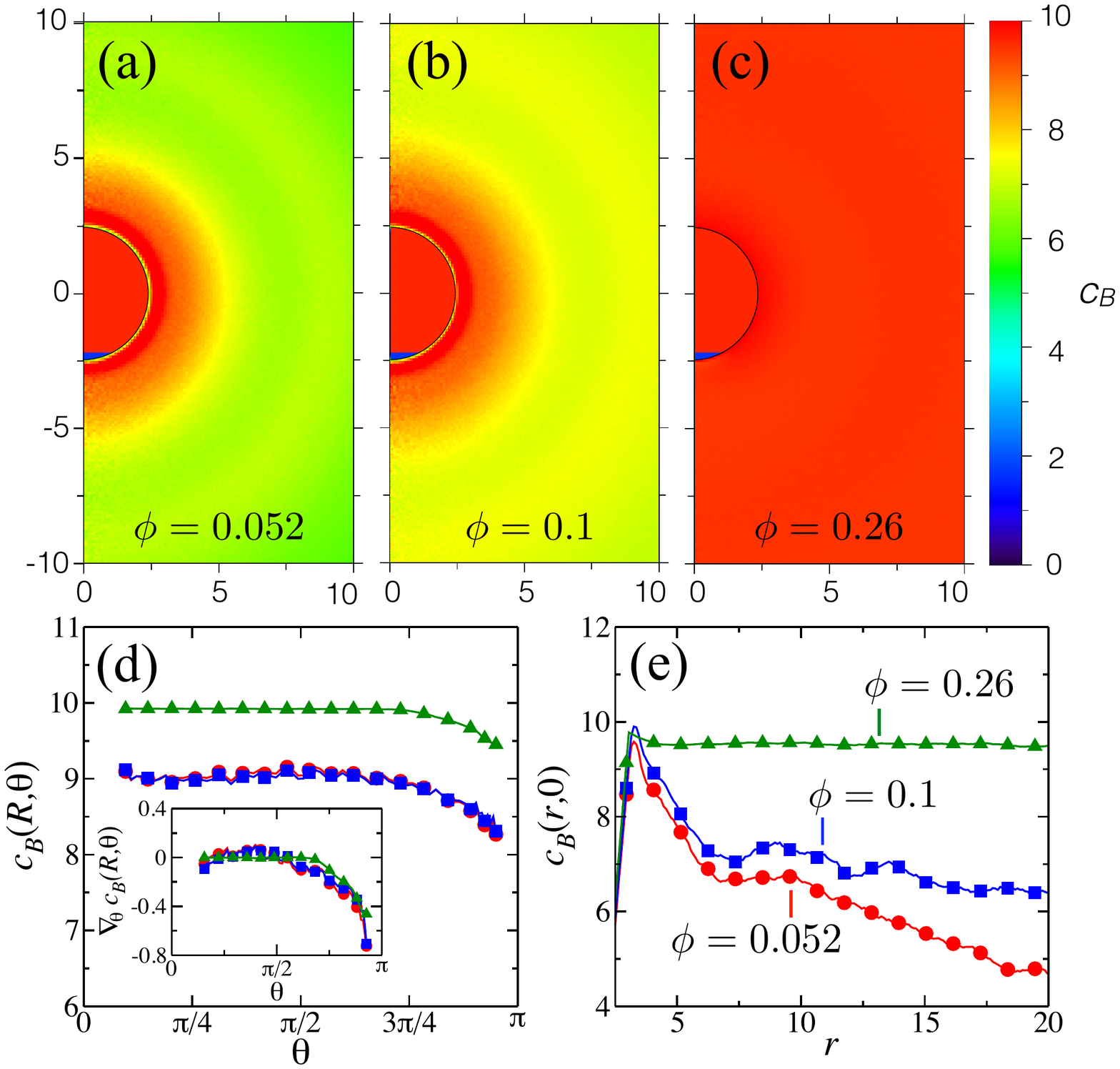}}
\caption{\label{fig:cB_theta_0_150} Steady-state product concentration fields, $c_B(r, \theta)$, in the vicinity of Janus motors with $\theta_C = 150^{\circ} $ obtained from simulations with motor volume fractions (a) $\phi = 0.052$, (b) $\phi = 0.1$ and (c) $\phi = 0.26$. The concentration fields in the tangential direction at the motor surface, $c_B(R,\theta)$, and in the radial direction along the motor axis, $c_B(r,0)$, are plotted in panels (d) and (e), respectively. The inset in (d) shows the gradient fields, $\nabla_{\theta} c_B(R,\theta)$, that are responsible for motor self-propulsion. In panels (d) and (e), the symbols are the labels of the simulation results with motor volume fractions $\phi = 0.052$ (red circles), $0.1$ (blue squares) and $0.26$ (green triangles).}
\end{figure}

Chemotactic interactions are controlled by gradients in the concentrations of reactive species. At large volume fractions of motors with large cap sizes one might anticipate that chemical gradients will be smoothed because of the high concentration of product molecules in the system, thereby reducing chemotactic interactions among the Janus motors. Indeed, as can be seen in the density maps of $c_B(r, \theta)$ in the steady state shown in Fig.~\ref{fig:cB_theta_0_150} (a-c). While significant gradients in the $B$ concentration can be seen in the radial direction at low volume fractions, a homogeneous distribution of $B$ particles is found for suspensions of large-cap motors at $\phi = 0.26$. These differences have only small effects on the self-propulsion speed of the motor. The propulsion speed is proportional to the tangential component of the concentration of the product at the surface of the motor.  As can be seen in Figure~\ref{fig:cB_theta_0_150} (d), the tangential component of the gradient at the motor surface at $r=R$ is effectively independent of the volume fraction for motors with large catalytic caps even though the average $B$ concentration increases substantially in the axial direction ($\theta = 0$) with volume fraction (see Fig.~\ref{fig:cB_theta_0_150}(e)). For dilute suspensions, a $B$-particle gradient field exists that is responsible for a chemotactic effect in which two motors are drawn close to one another due to the production of high local densities of the less repulsive $B$ particles. As the motor volume fraction increases, the density of $B$ in the radial direction increases and becomes more uniform.  As a result, only small inhomogeneities can be created locally around the catalytic surfaces of the motors and only weak chemotactic attractive interactions can be induced. Consequently, cluster formation is discouraged as the volume fraction of the large-cap motors in the suspension increases, as observed in Fig.~\ref{fig:g_r_with_chemo}.

\subsection{Hydrodynamic effects in the collective dynamics}
\label{sec:HI}
Hydrodynamic interactions among active swimmers have been investigated extensively~\cite{lauga2009,Baskaran_Marchetti_09,saintillan2012,Spagnolie_Lauga_12,Marchetti_etal_13,Bricard_etal_13,Zottl_Stark_14,Blaschke_etal_16}.  ``Puller" swimmers, characterized by an incoming fluid flow along the poles of the swimmer and outgoing flows in lateral directions, bring swimmers together along the direction of their swimming axis and are repulsive in the perpendicular directions, while the reverse behavior is observed for ``pusher" swimmers. (A discussion of the near- and far-field fluid velocity fields for the hard Janus model with $\theta_C = 90^{\circ}$ was given earlier~\cite{Reigh_etal_16}). As discussed in Sec.~\ref{sec:micro_model}, the characteristics of the near-field fluid flow patterns for chemically-powered Janus motors depend strongly on the size of the catalytic cap. In going from small to large cap sizes, the Janus motor may be classified in terms of the near-field flows as a puller swimmer for small caps ($\theta_C = 30^{\circ}$), a neutral swimmer for intermediate-size caps ($\theta_C = 90^{\circ}$), and a pusher swimmer for large-size caps ($\theta_C = 150^{\circ}$). The collective dynamics of suspensions of Janus motors may be affected by hydrodynamic interactions arising from these fluid flows, particularly for high volume fractions where the mean distance between motors is small.

\begin{figure}[htbp]
\centering
\resizebox{1.0\columnwidth}{!}{%
\includegraphics{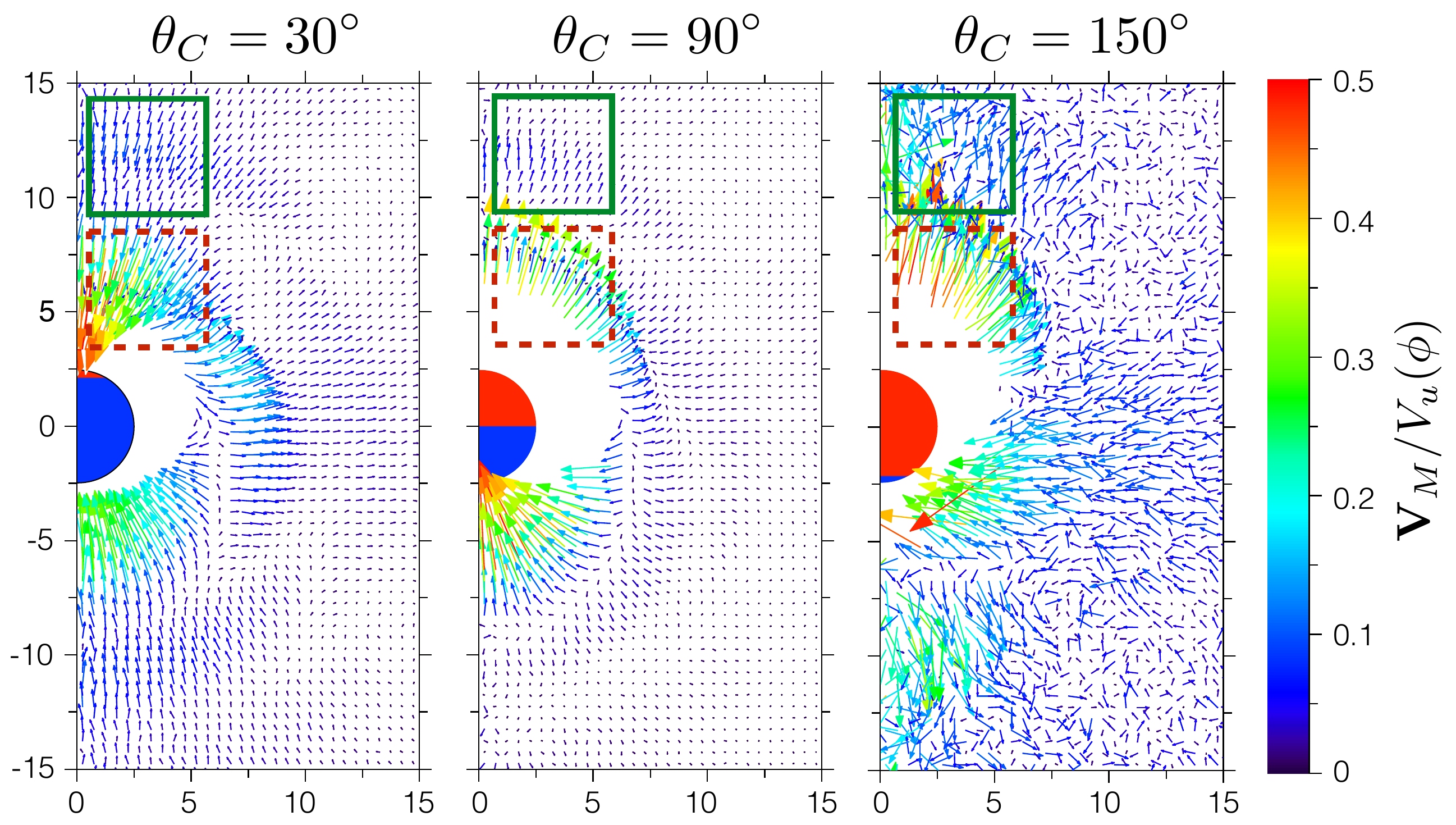}}
\caption{\label{fig:motor_v_field}Motor velocity fields, $\mathbf{V}_M(r,\theta)$, normalized by the average motor propulsion velocity, $\overline{\mbox{V}\raisebox{3mm}{}}_u(\phi)$, for various catalytic cap sizes at volume fraction $\phi = 0.1$. The regions of nearest and next-nearest neighbors from the motor surface are indicated by red-dashed and green squares, respectively.}
\end{figure}

The {\it motor} velocity fields in the suspension provide information on the effects of hydrodynamic flows on the collective motion. The vector field $\mathbf{V}_M$ for the axially-symmetric system is shown in Fig.~\ref{fig:motor_v_field} where the local values of the average velocity vector are represented as arrows for a suspension of volume fraction $\phi = 0.1$ (see Appendix\ref{app:mvec} for details). Of particular interest are the motor flows in the region in front (i.e, at the catalytic head) of the motor where the $A \to B$ reactions take place and give rise to the strongest concentration-mediated chemotactic interaction. In  Fig.~\ref{fig:motor_v_field} the red-dashed and green-solid squares highlight regions around the nearest ($r\simeq 6$) and next-nearest neighbors ($r\simeq 12$), respectively.  These distances correspond to the locations of the peaks in the plots of the radial distribution function $g(r)$ shown in Fig.~\ref{fig:g_r_with_chemo}. Within the red-dashed regions, one sees that the average motor flow velocity points inward toward the surface of the motor for small cap sizes, while for large cap sizes the motor flow velocities are outgoing from the motor surface.

\begin{figure}[htbp]
\centering
\resizebox{0.8\columnwidth}{!}{%
\includegraphics{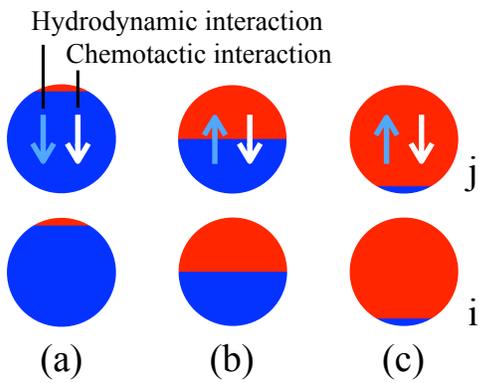}}
\caption{\label{fig:chemo_hydro_interaction}Aligned configurations of Janus motors with (a) $\theta_C = 30^{\circ}$, (b) $\theta_C = 90^{\circ}$ and (c) $\theta_C = 150^{\circ}$, where the light blue and white arrows are the direction of motion of motor $j$ induced by the motion of motor $i$ through the hydrodynamic and chemotactic interactions, respectively.}
\end{figure}

Apart from direct intermolecular forces, the net interaction between motors is determined by chemotactic and hydrodynamic contributions. As illustrated in Fig.~\ref{fig:chemo_hydro_interaction}, while the former interaction is attractive for all cap sizes, the hydrodynamic interaction is attractive in the region directly ahead of the motor for $\theta_C = 30^{\circ}$ and repulsive for $\theta_C = 90^{\circ}$ and $150^{\circ}$. For $\theta_C = 30^{\circ}$, the fact that both interactions are attractive gives rise to the incoming motor flows seen in Fig.~\ref{fig:motor_v_field}. By contrast, for $\theta_C = 90^{\circ}$, the repulsive hydrodynamic interaction dominates over chemotactic attraction resulting in outgoing motor flows discussed earlier. These results suggest that hydrodynamic interactions may either enhance or reduce attractive interactions for the Janus motors with $\theta_C = 30^{\circ}$ and $90^{\circ}$, respectively. The interplay of chemotactic and hydrodynamic interactions explains the enhanced clustering observed in the Janus motors with small cap sizes. Since the dynamics of Janus motors is largely diffusive for suspensions of Janus motors with large cap sizes, the clustering mechanism is primarily influenced by the chemotactic interactions discussed in the previous section and hydrodynamic interactions play only a minor role. Nevertheless, hydrodynamic effects can been seen by comparing the solvent and motor velocity fields depicted in Figs.~\ref{fig:cB_solvent_flow} and \ref{fig:motor_v_field}, where it is evident that these two fields have the same characteristic flow patterns.

\begin{figure}[htbp]
\centering
\resizebox{1.0\columnwidth}{!}{%
\includegraphics{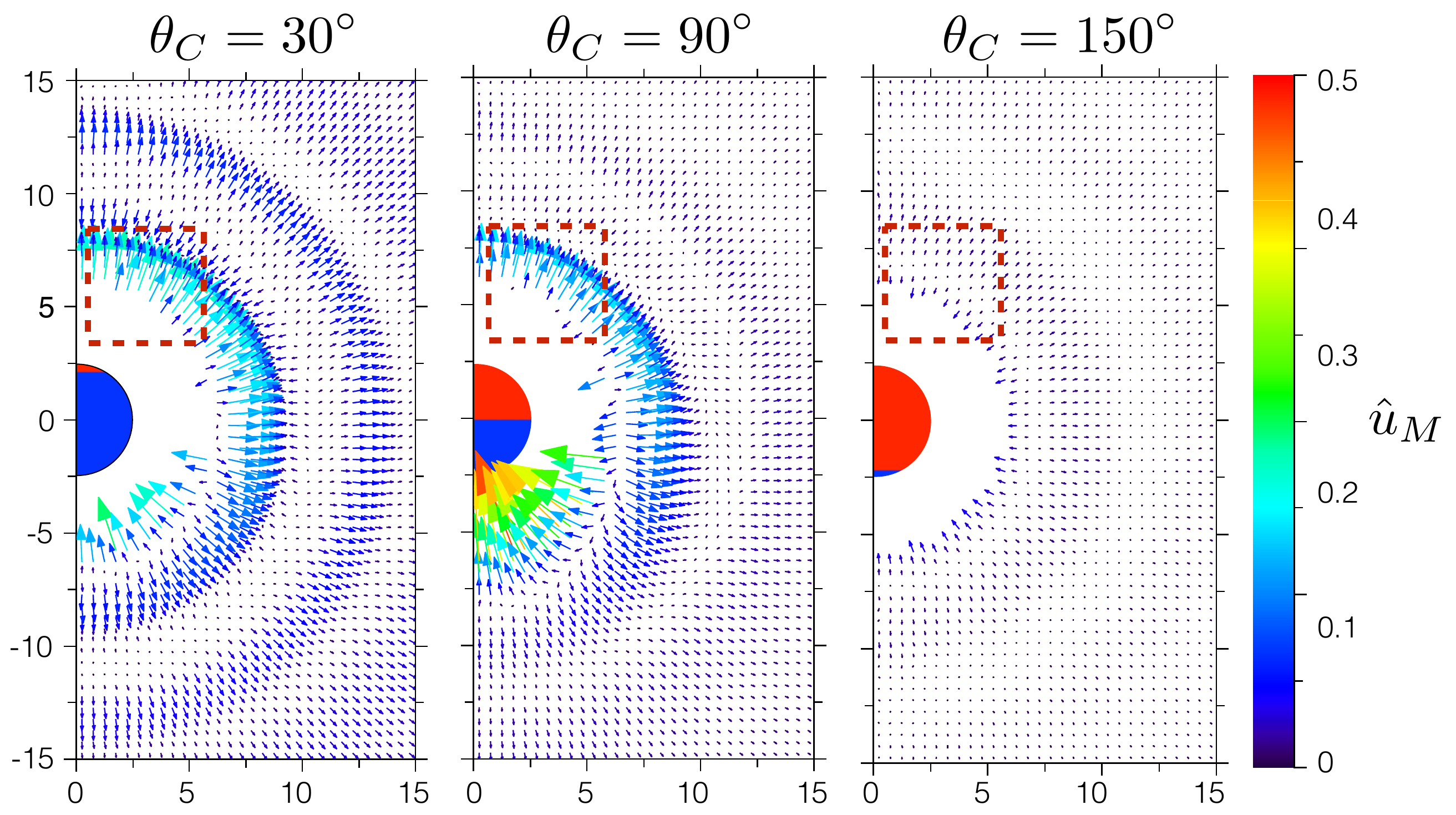}}
\caption{\label{fig:motor_ori_field}Motor orientation fields, $\hat{{\bf u}}_M(r,\theta)$, for various cap sizes at volume fraction $\phi = 0.1$. The nearest-neighbor regions are indicated by red-dashed squares.}
\end{figure}

To examine the degree of orientational ordering that accompanies the correlations in the motor velocity field for $\theta_C = 30^{\circ}$ discussed above, we computed the average of the relative orientation of a motor, $\hat{{\bf u}}_M$ (see Appendix \ref{app:mvec} for details). This field is plotted in Fig.~\ref{fig:motor_ori_field}. At the nearest neighbor distance in the front of the motor (highlighted by a red-dashed square), the average orientation is in the same direction as that of the motor at the origin. Similar orientational ordering can be seen for $\theta_C = 90^{\circ}$; however, for $\theta_C = 150^{\circ}$ neighboring motors are less ordered due to the dominance of diffusive motion and lack of active motion for this cap size.
These results suggest that the clustering mechanism for self-diffusiophoretic Janus motors is not due primarily to motility-induced phase separation~\cite{cates2013}, where motors are dynamically jammed due to head-on collisions, but instead arises from attractive hydrodynamic and chemotactic interactions.

\subsection{Elimination of chemotactic interactions}

Since all Janus motors in a suspension at a given volume fraction are identical, product molecules produced on one motor may interact with other motors giving rise to the chemotactic interactions discussed above. Of course, the catalytic reactions are also essential for the self-generated chemical gradients that are responsible for single motor propulsion. In order to gauge the relative importance of chemotactic and hydrodynamic effects we must be able to selectively turn off these interactions while not disturbing the self-propulsion diffusiophoretic mechanism for single motors which involves both local concentration gradients and coupling to fluid flow.

\label{sec:ECI}
\begin{figure}[htbp]
\centering
\resizebox{1.0\columnwidth}{!}{%
\includegraphics{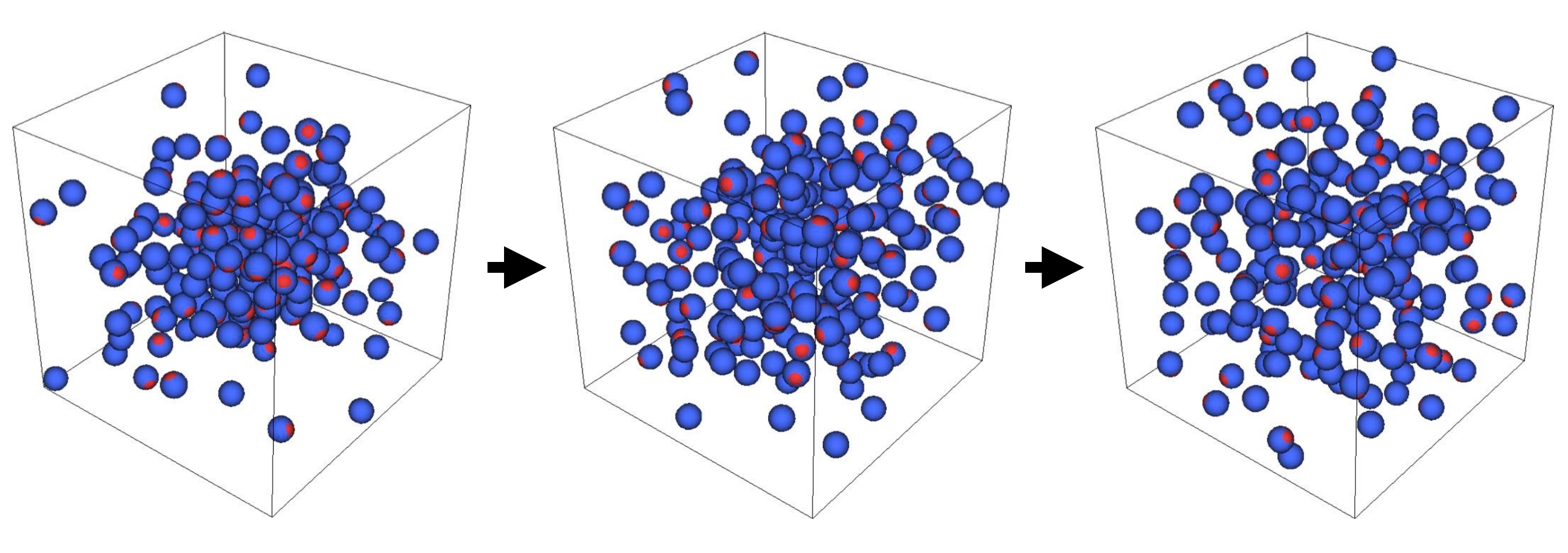}}
\caption{\label{fig:no_CI_snapshot} Starting from a cluster of Janus motors small cap sizes, in the absence of chemotactic interactions the cluster gradually breaks apart. (Movie online)}
\end{figure}

This can be achieved in the following way: we consider a collection of Janus motors in which each motor $i$ produces a distinct product $B_i$ that interacts with motor $i$ as a product particle but as a non-reactive $A$ particle with all other motors $j$. In this way only the self-generated concentration gradient of $B_i$ is responsible for the propulsion of that motor. In this model the concentration-mediated chemotactic attraction is turned off while hydrodynamic interactions between Janus motors resulting from self-propulsion remain. (This method of turning off chemotactic interactions was used in an earlier investigation of the collective dynamics of oligomeric motors on a filament~\cite{huang_kapral_16}. Starting from an initial cluster configuration, a time sequence of motor configurations in which the system evolves in the absence of chemotactic interactions is shown in Fig.~\ref{fig:no_CI_snapshot}. The cluster of Janus motors gradually breaks apart and the system reaches a steady state in which the motors are homogeneously distributed beyond the first solvation shell.

\begin{figure}[htbp]
\centering
\resizebox{0.9\columnwidth}{!}{%
\includegraphics{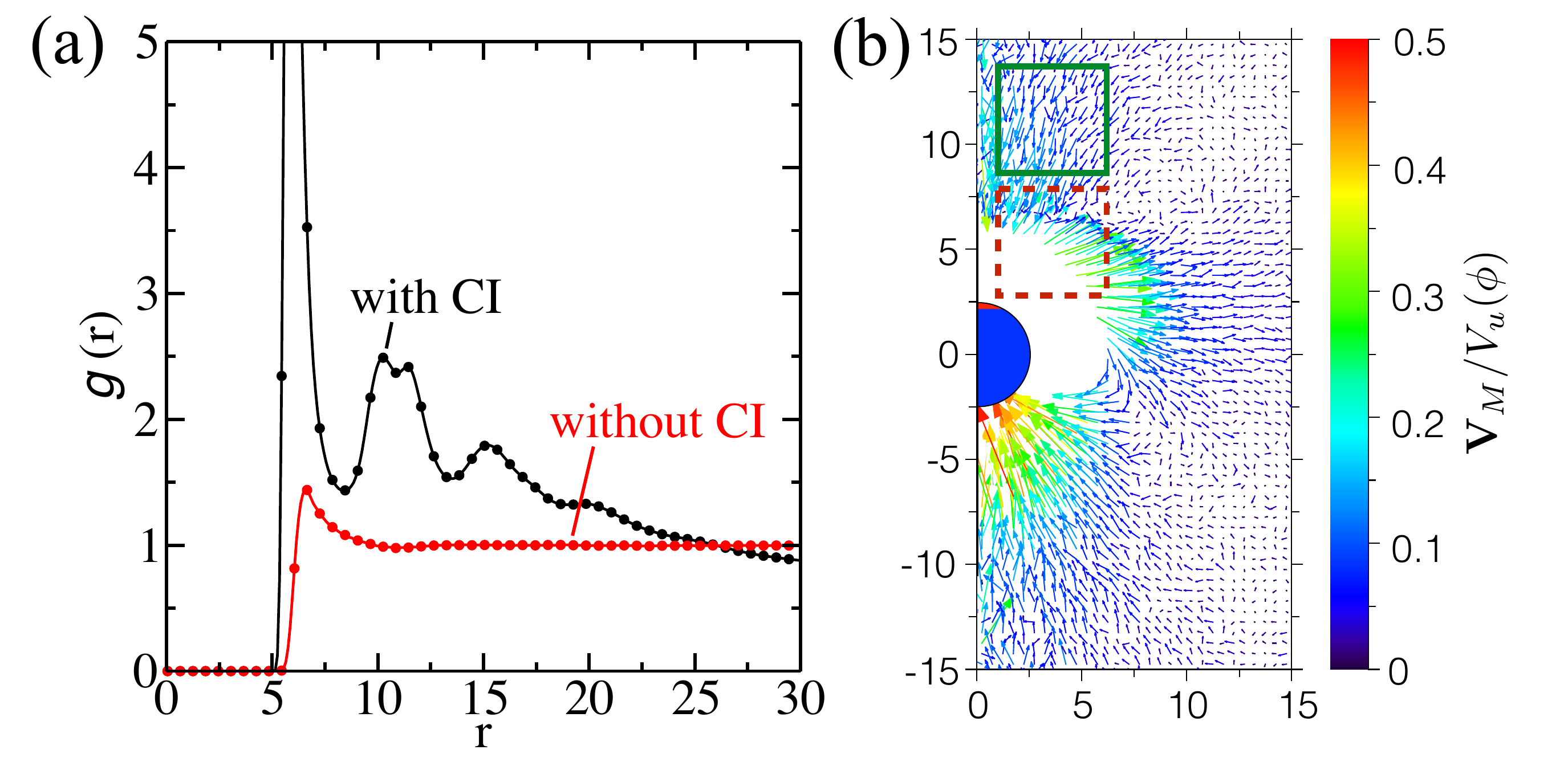}}
\caption{\label{fig:no_CI} (a) The radial distribution functions for the Janus motors with $\theta_C = 30^{\circ}$ in the presence and absence of chemotactic interactions. (b) The motor velocity field in the absence of chemotactic interactions. The red-dashed and green squares indicate nearest and next-nearest neighbor regions.}
\end{figure}

From an examination of the steady state radial distribution functions in Fig.~\ref{fig:no_CI} (a), the disappearance of clustering in the absence of concentration-mediated interactions is evident by the lack of structure in $g(r)$.  Nonetheless Janus motors still interact through hydrodynamic interactions in the absence of chemotactic attractions. Comparing Figs.~\ref{fig:motor_v_field} and \ref{fig:no_CI} (b) for $\theta_C = 30^{\circ}$, it is clear that turning off chemotactic interactions has a significant impact on the motor velocity fields at short distances but a weaker influence at long distances. These results suggest that long-ranged interactions are mediated by hydrodynamic interactions, which bring Janus motors toward each other, while the short-ranged interactions are primarily due to chemotactic effects.

\subsection{Screening due to bulk reactions}
\label{sec:bulk}
\begin{figure}[htbp]
\centering
\resizebox{1.0\columnwidth}{!}{%
\includegraphics{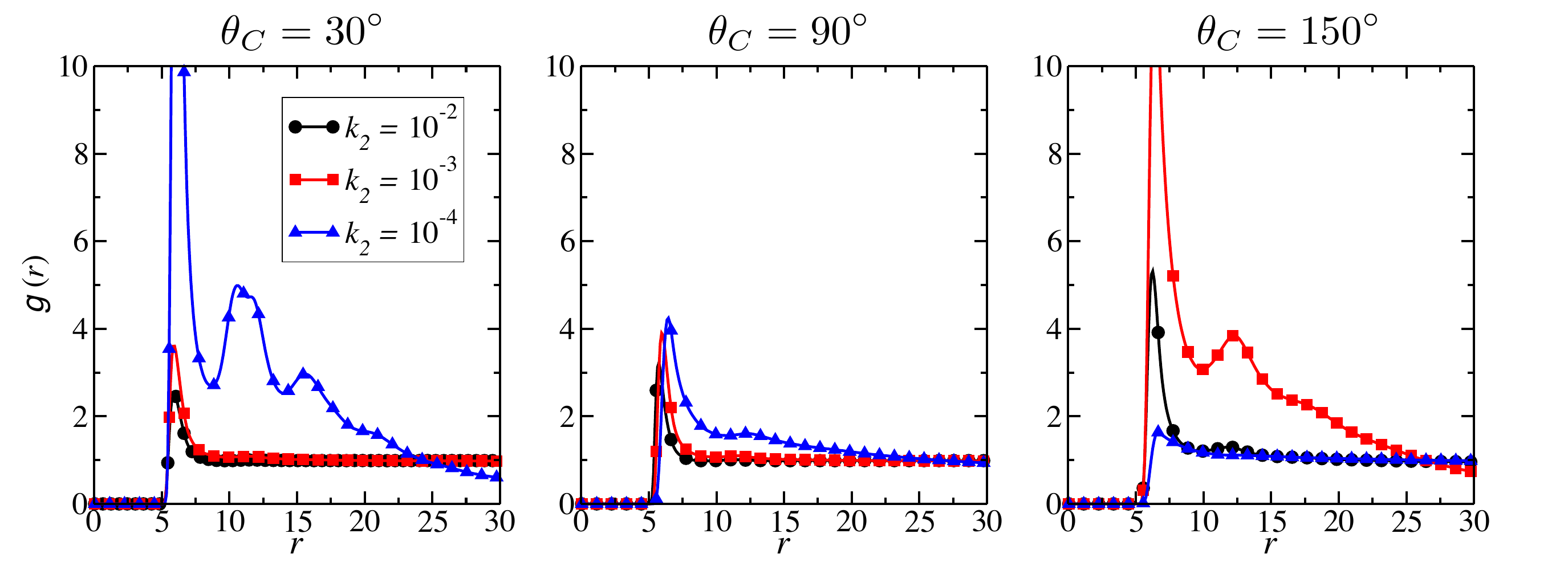}}
\caption{\label{fig:g_r_k2} Radial distribution functions, $g(r)$, of Janus motors with small (left panel), medium (middle panel) and large (right panel) catalytic surface in systems at volume fraction $\phi = 0.052$ with various bulk reaction rate constants, $k_2 = 10^{-2}$ (black circles), $10^{-3}$ (red squares) and $10^{-4}$ (blue triangles).}
\end{figure}

The bulk reaction $B \stackrel{k_2}{\rightarrow }A$ is not only responsible for the supply of fuel that maintains the system in a nonequilibrium state but it also has an effect on the collective dynamics. Consider a single motor in solution. In the absence of a bulk reaction with fuel supplied only at large distances from the motor the $c_B(r,\theta)$ steady-state concentration field produced at the surface of the motor by the catalytic reaction $A \to B$ will decay in the radial direction as $c_B(r,\theta) \sim 1/r$~\cite{kapral2013}. When there is a bulk reaction this field decays as $c_B(r,\theta) \sim e^{-\kappa r}/r$~\cite{Huang_etal_16}. Consequently, the concentration fields are ``screened" by the bulk reaction, and the screening length $\kappa^{-1}$ will determine the range of chemotactic interactions among motors. A slow back reaction with a small value of $k_2$ will lead to longer-ranged chemotactic interactions that may influence motor clustering.

To study this effect, we consider a suspension of motors with various cap sizes and volume fraction $\phi=0.052$ in bulk reacting media with different rate constants $k_2$. In Fig.~\ref{fig:g_r_k2}, the radial distribution function of the Janus motors is shown for $k_2 = 10^{-2}$ (black circles), $10^{-3}$ (red squares) and $10^{-4}$ (blue triangles), corresponding screening lengths $\kappa^{-1} \approx 2.6$, $8.4$ and $26.45$, respectively. For the fastest bulk reaction where $k_2 = 0.01$, the concentration-mediated interaction is strongly screened resulting in virtually no clustering for all cap sizes.  As the bulk reaction rate decreases, clustering is found for Janus motors with small and intermediate cap sizes ($\theta_C \leq 90^{\circ}$) as can be seen by the increase in $g(r)$ beyond the first peak. Note also that due to the different characteristics of hydrodynamic interactions for different cap sizes discussed earlier, Janus motors with a small cap size ($\theta_C = 30^{\circ}$) aggregate more strongly than those with $\theta_C = 90^{\circ}$. Interestingly, for the large cap size, $\theta_C = 150^{\circ}$, there is an intermediate range of $k_2$ values where stronger motor clustering is observed. This is due to the fact that the chemotactic interaction between motors is more strongly screened when $k_2$ is large and motors move independently without clustering, while for small $k_2$ where the screening length is large, gradients of chemical species are smoothed since a large fraction of solvent particles consists of uniformly distributed product. This reduction in clustering due to high local uniform concentrations of product is similar to the reduction of clustering observed for motors with large-size caps for high volume fractions discussed in Sec.~\ref{sec:150}.

\section{Summary}\label{sec:conclusion}

Since the propulsion of a Janus motor depends on the asymmetric chemical reactions on its catalytic cap, changes the cap area have a significant influence on the concentration and flow fields near the motor surface and in the surrounding fluid. These changes are reflected in the chemotactic and hydrodynamic interactions that determine the collective behavior of suspensions of active Janus motors. Using particle-based simulations of Janus motor suspensions it was shown that chemotactic interactions are the dominant factor in motor cluster formation. Depending on the catalytic cap size, hydrodynamic interactions can either enhance or decrease the tendency to cluster by chemotactic interactions. Since chemotactic interactions can be turned off in our simulations, the relative importance of chemotactic and hydrodynamic effects can be determined without modifying the diffusiophoretic propulsion mechanism that underlies the phenomena. In the simulations the system was driven out of equilibrium by bulk phase reactions that remove product and add fuel particles. As a result the concentration fields decay more rapidly with distance than if fuel were supplied at the boundaries. Consequently, the chemotactic interaction between motors is screened, and decreasing the bulk reaction rate constant and thereby increasing the screening length enhances clustering in suspensions provided the catalytic cap size is not too large. Thus, the way in which the system is maintained in a nonequilibrium state is an important factor to consider when studying collective dynamics, especially for three-dimensional systems when fuel is supplied at the boundaries.

\section*{Acknowledgements}
This work was supported by grants from the Natural Sciences and Engineering Research Council of Canada and Compute Canada.
\appendix

\section{Simulation algorithm and parameters}\label{app:sim}We use a simulation scheme developed previously, which combines molecular dynamics with multiparticle particle collision dynamics~\cite{Malevanets_Kapral_00}. Each Janus motor is modeled as a hard sphere and solvent $A$ and $B$ particles undergo reactive and nonreactive collision with it~\cite{Huang_etal_16} as discussed in Sec.~\ref{sec:micro_model}. The direct interaction between two Janus motors is modeled by a repulsive Lennard-Jones potential. The dynamics of solvent particles is carried out by reactive multiparticle collision dynamics~\cite{Rohlf_Fraser_Kapral_08}, where non-interacting fluid particles evolve in the presence of the interacting motors and experience effective collisions at discrete time intervals $\tau$. In the collision steps, solvent particles are sorted into a grid of cube cells of linear size $\sigma$, where effective multiparticle collisions and reactions $B \to A$  with reaction rate constant $k_2$ take place independently in each cell.

The system consists of a collection of $N_J$ Janus motors and $N_S = N_A + N_B$ point $A$ and $B$ particles, each of mass $m$, confined to a periodic cubic box with linear size $L$ and volume $V=L^3$ . The Janus motor is modeled as a sphere of radius $R$ with volume $V_J = \frac{4}{3} \pi R^3$, mass $M = c_0 V_J m$ and moment of inertia $I =\frac{2}{5}MR^2$, where the total number density of fluid particles is $c_0 = N_S/(V - N_J V_J)$. Solvent particles interact with the Janus motor through hard potentials while  a repulsive Lennard-Jones potential is used for interactions between two Janus motors. The Lennard-Jones potential is $V(r) = 4 \epsilon [(\sigma_J/r)^{12} - (\sigma_J/r)^6 + 0.25]$, which vanishes at the cutoff distance $r_c = 2^{1/6}\:\sigma_J$. The system is maintained in a nonequilibrium steady state by using reactive multiparticle collision dynamics~\cite{Rohlf_Fraser_Kapral_08}, where the reaction $B\stackrel{k_2}{\rightarrow} A$ takes place in the bulk solution. Grid-shifting is employed to ensure Galilean invariance~\cite{Ihle_Kroll_01}. The time evolution of the entire system is carried out using a hybrid MD-MPCD scheme~\cite{Malevanets_Kapral_00}.

The results are reported in dimensionless units where mass is in units of $m$, length in units of $\sigma$, energies in units of $k_BT$ and time in units of $t_0 = \sqrt{m \sigma^2/k_BT}$. In these units, we have $L = 60$, $R = 2.5$, $M = 655$, and $I = 1636$. The multiparticle collision time was set to $\tau = 0.1$. Between two consecutive multiparticle collisions, solvent particles propagate freely with $\mathbf{r}(t+\delta t) = \mathbf{r}(t) + \mathbf{v}(t) \delta t$, where $\delta t = 0.01$ is the molecular dynamics time step size. For the simulation conditions above the fluid viscosity is $\eta=7.93$ and the common self-diffusion coefficient of the $A$ and $B$  species is $D=0.07$. The Lennard-Jones parameters are $\epsilon = 1$ and $\sigma_J = 6$.
For a solvent density $c_0 = 10$, motor velocity $V_u \leq 0.02$ and motor radius of $R = 2.5$, the kinematic viscosity of the fluid is $\nu = \eta/c_0 = 0.793$, the P\'{e}clet number is Pe $= V_u R / D < 0.7$, the Reynolds number is Re$ = V_u R/\nu < 0.06$, and the Schmidt number is $S_c = \nu / D \simeq 11$.

\section{Vector fields of Janus motors}\label{app:mvec}
The average vector field of a physical quantity, $\mathbf{W}_M$, in the vicinity of a motor is determined by measuring the average field in a coordinate frame defined by two in-plane vectors $\hat{\mathbf{n}}_1$ and $\hat{\mathbf{n}}_2$. We consider the  motor velocity and orientational fields, $\mathbf{W}_M = \mathbf{V}_M$  or $\hat{\mathbf{u}}_M$, respectively. The vector field can be written in terms of the in-plane coordinates $\mathbf{W}_M = \mathbf{W}_M(\mathbf{r}_S)$, where $\mathbf{r}_S$ denotes the in-plane vector. If the orientation vector $\hat{\mathbf{u}}_i$ of motor $i$ is chosen to be $\hat{\mathbf{n}}_1$, the vector $\hat{\mathbf{n}}_2$ is given by
\begin{equation}
\hat{\mathbf{n}}_2 = \frac{\mathbf{r}_{ij} - \mathbf{r}_{ij}\cdot \hat{\mathbf{n}}_1 }{|\mathbf{r}_{ij} - \mathbf{r}_{ij}\cdot \hat{\mathbf{n}}_1|}.
\end{equation}
By using the position of the motor $i$ as the origin, the vector pointing from the origin to the motor $j$ is $\mathbf{r}_{ij} = \mathbf{r}_{j} - \mathbf{r}_{i}$. The coordinates of the motor $j$ projected on the plane can be computed by
\begin{equation}
\mathbf{r}_S = (\mathbf{r}_{ij}\cdot \hat{\mathbf{n}}_1)\hat{\mathbf{n}}_1 + (\mathbf{r}_{ij}\cdot \hat{\mathbf{n}}_2)\hat{\mathbf{n}}_2,
\end{equation}
and the vector of a physical quantity $\mathbf{w}$ of motor $j$ at the position $\mathbf{r}_S$ can be expresses as
\begin{equation}
\mathbf{w}_{S}(\mathbf{r}_S) = (\mathbf{w}_j\cdot \hat{\mathbf{n}}_1)\hat{\mathbf{n}}_1 + (\mathbf{w}_j\cdot \hat{\mathbf{n}}_2)\hat{\mathbf{n}}_2.
\end{equation}
Thus, by sorting the neighboring motors into a square lattice with linear size of $0.5$ according to the in-plane coordinates $\mathbf{r}_S$ the average vector field, $\mathbf{W}_M$, in each lattice cell is given by the average over the ensemble of motors and over time.


\begin{thebibliography}{10}
\expandafter\ifx\csname url\endcsname\relax
  \def\url#1{{\tt #1}}\fi
\expandafter\ifx\csname urlprefix\endcsname\relax\def\urlprefix{URL }\fi
\providecommand{\eprint}[2][]{\url{#2}}

\bibitem{alberts-cell}
Alberts B, Bray D, Lewis J, Raff M, Roberts K and Watson J~D 2002 {\em
  Molecular Biology of the Cell\/} 3rd ed (Garland Science)

\bibitem{lauga2009}
Lauga E and Powers T~R 2009 {\em Rep. Prog. Phys.\/} {\bf 72} 096601

\bibitem{Baskaran_Marchetti_09}
Baskaran A and Marchetti M~C 2009 {\em Proc. Natl. Acad. Sci. USA\/} {\bf 106}
  15567

\bibitem{saintillan2012}
Saintillan D and Shelley M~J 2012 {\em J. Roy. Soc. Interface\/} {\bf 9}
  571--585

\bibitem{Spagnolie_Lauga_12}
Spagnolie S~E and Lauga E 2016 {\em J. Fluid Mech.\/} {\bf 700} 105

\bibitem{Marchetti_etal_13}
Marchetti M~C, Joanny J~F, Ramaswamy S, Liverpool T~T, Prost J, Rao M and Simha
  R~A 2013 {\em Rev. Mod. Phys.\/} {\bf 85} 1143

\bibitem{Bricard_etal_13}
Bricard A, Caussin J~B, Desreumaux N, Dauchot O and Bartolo D 2013 {\em
  Nature\/} {\bf 503} 95

\bibitem{Zottl_Stark_14}
Z{\"{o}}ttl A and Stark H 2014 {\em Phys. Rev. Lett.\/} {\bf 112} 118101

\bibitem{Blaschke_etal_16}
Blaschke J, Maurer M, Memon K, Z{\"{o}}ttl A and Stark H 2016 {\em Soft
  Matter\/} {\bf 12} 9821

\bibitem{vicsek1995}
Vicsek T, Czir{\'o}k A, Ben-Jacob E, Cohen I and Shochet O 1995 {\em Phys. Rev.
  Lett.\/} {\bf 75}(6) 1226--1229

\bibitem{chate2008}
Chat{\'e} H, Ginelli F, Gr{\'e}goire G and Raynaud F 2008 {\em Phys. Rev. E\/}
  {\bf 77}(4) 046113

\bibitem{peruani2006}
Peruani F, Deutsch A and B{\"a}r M 2006 {\em Phys. Rev. E\/} {\bf 74}(3) 030904

\bibitem{redner2013}
Redner G~S, Hagan M~F and Baskaran A 2013 {\em Phys. Rev. Lett.\/} {\bf 110}
  055701

\bibitem{cates2013}
Cates M~E and Tailleur J 2013 {\em EPL\/} {\bf 101} 20010

\bibitem{bialke2013}
Bialk{\'e} J, L{\"o}wen H and Speck T 2013 {\em EPL\/} {\bf 103} 30008

\bibitem{Palacci_eta_13}
Palacci J, Sacanna S, Steinberg A~P, Pine D~J and Chaikin P~M 2013 {\em
  Scinece\/} {\bf 339} 936

\bibitem{wysocki2014}
Wysocki A, Winkler R~G and Gompper G 2014 {\em EPL\/} {\bf 105} 48004

\bibitem{takatori2015}
Takatori S~C and Brady J~F 2015 {\em Phys. Rev. E\/} {\bf 91}(3) 032117

\bibitem{speck2015}
Speck T, Menzel A~M, Bialk{\'e} J and L{\"o}wen H 2015 {\em J. Chem. Phys.\/}
  {\bf 142} 224109

\bibitem{Zottl_Stark_16}
Z{\"{o}}ttl A and Stark H 2016 {\em J. Phys.: Condens. Matter\/} {\bf 28}
  253001

\bibitem{SenRev:13}
Wang W, Duan W, Ahmed S, Mallouk T~E and Sen A 2013 {\em Nano Today\/} {\bf 8}
  531 -- 554

\bibitem{wangbook:13}
Wang J 2013 {\em Nanomachines: Fundamentals and Applications\/} (Weinheim:
  Wiley-VCH)

\bibitem{kapral2013}
Kapral R 2013 {\em J. Chem. Phys.\/} {\bf 138} 020901

\bibitem{Ma_Hahn_Sanchez_15}
Ma X, Hahn K and Sanchez S 2015 {\em J. Am. Chem. Soc.\/} {\bf 137} 4976

\bibitem{Jiang_etal_10}
Jiang S, Chen Q, Tripathy M, Luijten E, Schweizer K~S and Granick S 2010 {\em
  Adv. Mater.\/} {\bf 22} 1060

\bibitem{ibele2009}
Ibele M, Mallouk T and Sen A 2009 {\em Angew. Chem. Int. Ed.\/} {\bf 48}
  3308--3312

\bibitem{Theurkauff_etal_12}
Theurkauff I, Cottin-Bizonne C, Palacci J, Ybert C and Bocquet L 2012 {\em
  Phys. Rev. Lett.\/} {\bf 108} 268303

\bibitem{Thakur_Kapral_12}
Thakur S and Kapral R 2012 {\em Phys. Rev. E\/} {\bf 85} 026121

\bibitem{Wang_etal_13}
Wang W, Duan W, Sen A and Mallouk T~E 2013 {\em Proc. Natl. Acad. Sci. USA\/}
  {\bf 110} 17744

\bibitem{Buttinoni_etal_13}
Buttinoni I, Bialk{\'{e}} J, L{\"{o}}wen H, Bechinger C and Speck T 2013 {\em
  Phys. Rev. Lett.\/} {\bf 110} 238301

\bibitem{kapral2014}
Colberg P~H, Reigh S~Y, Robertson B and Kapral R 2014 {\em Acc. Chem. Res.\/}
  {\bf 47} 3504

\bibitem{Pohl_Stark_14}
Pohl O and Stark H 2014 {\em Phys. Rev. Lett.\/} {\bf 112} 238303

\bibitem{Saha_etal_14}
Saha S, Golestanian R and Ramaswamy S 2014 {\em Phys. Rev. E\/} {\bf 89} 062316

\bibitem{Pohl_Stark_15}
Pohl O and Stark H 2015 {\em Eur. Phys. J. E\/} {\bf 38} 93

\bibitem{wang2015}
Wang W, Duan W, Ahmed S, Sen A and Mallouk T~E 2015 {\em Acc. Chem. Res.\/}
  {\bf 48} 1938--1946

\bibitem{huang_kapral_16}
Huang M~J and Kapral R 2016 {\em Eur. Phys. J. E\/} {\bf 39} 36

\bibitem{Colberg_Kapral_17}
Colberg P and Kapral R 2017 {\em arXiv:1703.03034\/}

\bibitem{anderson1989}
Anderson J~L 1989 {\em Ann. Rev. Fluid Mech.\/} {\bf 21} 61--99

\bibitem{golestanian2005}
Golestanian R, Liverpool T~B and Ajdari A 2005 {\em Phys. Rev. Lett.\/} {\bf
  94} 220801

\bibitem{Popescu_etal_10}
Popescu M~N, Dietrich S, Tasinkevych M and Ralston J 2010 {\em Eur. Phys. J.
  E\/} {\bf 31} 351

\bibitem{Popsescu_etal_16}
Popescu M~N, Uspal W~E and Dietrich S 2016 {\em Eur. Phys. J. Special Topics\/}
  {\bf 225} 2189

\bibitem{Mood_etal_09}
Sharifi-Mood N, Mozaffari A and C{\'{o}}rdova-Figueroa U~M 2016 {\em J. Fluid
  Mech.\/} {\bf 798} 910

\bibitem{Huang_etal_16}
Huang M~J, Schofield J and Kapral R 2016 {\em Soft Matter\/} {\bf 12} 5581

\bibitem{Malevanets_Kapral_99}
Malevanets A and Kapral R 1999 {\em J. Chem. Phys.\/} {\bf 110} 8605

\bibitem{Malevanets_Kapral_00}
Malevanets A and Kapral R 2000 {\em J. Chem. Phys.\/} {\bf 112} 72609

\bibitem{kapral:08}
Kapral R 2008 {\em Adv. Chem. Phys.\/} {\bf 140} 89--146

\bibitem{gompper:09}
Gompper G, Ihle T, Kroll D~M and Winkler R~G 2009 {\em Adv. Polym. Sci.\/} {\bf
  221} 1--87

\bibitem{anderson:83}
Anderson J~L 1983 {\em Phys. Fluids\/} {\bf 26} 2871

\bibitem{Reigh_etal_16}
Reigh S~Y, Huang M~J, Schofield J and Kapral R 2016 {\em Phil. Trans. R. Soc.
  A\/} {\bf 374} 20160140

\bibitem{Julicher_Prost_09}
J\"ulicher F and Prost J 2009 {\em Eur. Phys. J. E\/} {\bf 29} 27--36

\bibitem{Huang_Schofield_Kapral_17}
Huang M~J, Schofield J and Kapral R 2017 {\em J. Phys. A: Math. Theor.\/} {\bf
  50} 074001

\bibitem{Rohlf_Fraser_Kapral_08}
Rohlf K, Fraser S and Kapral R 2008 {\em Computs. Phys. Commun.\/} {\bf 179}
  132

\bibitem{Ihle_Kroll_01}
Ihle T and Kroll D~M 2001 {\em Phys. Rev. E\/} {\bf 63} 020201

\end{thebibliography}
\providecommand{\noopsort}[1]{}\providecommand{\singleletter}[1]{#1}
\providecommand{\newblock}{}

\end{document}